\documentclass{article}

\usepackage{fancyhdr}
\usepackage[english]{babel}
\usepackage{amsmath,amssymb,amsthm} 
\usepackage{hyperref}   
\usepackage{braket}
\usepackage{cite}
\usepackage{enumitem}
\usepackage{bm}

\theoremstyle{plain}
\newtheorem{theorem}{Theorem}[section]
\newtheorem{lemma}[theorem]{Lemma}

\newtheorem{proposition}[theorem]{Proposition}

\theoremstyle{definition}
\newtheorem{definition}[theorem]{Definition}

\newtheorem{observation}[theorem]{Observation}

\theoremstyle{remark}
\newtheorem{remark}[theorem]{Remark}

\theoremstyle{plain}
\newtheorem*{result}{Main Theorem}

\title{Nontrivial Riemann Zeros as Spectrum}
\author{Enderalp Yakaboylu\thanks{yakaboylu@proton.me}}

\begin{document}





\maketitle

\begin{abstract}
We present the Riemann operator $ \hat{\mathcal{R}} $, a densely defined non-symmetric operator acting on $ L^2([0,\infty)) $ whose point spectrum is given by the nontrivial zeros of the Riemann zeta function $ \zeta(s) $:
\[
\sigma_{\mathrm{p}}(\hat{\mathcal{R}})=\left\{i\left(\rho-1/2\right)\mid\rho\in\mathcal{Z}_\zeta\right\}, \qquad \mathcal{Z}_\zeta:=\left\{\rho\mid\zeta(\rho)=0,\;0<\Re(\rho)<1\right\} \, .
\]
We prove the existence of a linear operator $ \hat{W} $ that intertwines the Riemann operator $ \hat{\mathcal{R}} $ with its Hilbert-space adjoint $ \hat{\mathcal{R}}^* $, i.e., $ \hat{\mathcal{R}}\,\hat{W}=\hat{W}\,\hat{\mathcal{R}}^* $. Strict positivity of the intertwiner $ \hat{W} $ is equivalent to the Riemann Hypothesis together with the simplicity of the nontrivial zeros:
\[
\hat{W}>0\quad\Longleftrightarrow\quad\Re(\rho)=1/2\;\text{ and }\;\zeta^{(1)}(\rho)\neq0\qquad\forall\,\rho\in\mathcal{Z}_\zeta \, .
\]
\end{abstract}




\section{Introduction}
\label{sec_intro}

For complex numbers \( s \) with \( \Re(s)>1 \), the Riemann zeta function is defined by its Dirichlet series, which factors as an Euler product over the primes \( \mathbb{P} \):
\[
  \zeta(s)=\sum_{n=1}^\infty\frac{1}{n^s}=\prod_{p\in\mathbb{P}}\frac{1}{1-p^{-s}} \, .
\]
Through analytic continuation, \( \zeta(s) \) extends to a meromorphic function on the entire complex plane \( \mathbb{C} \) with a single simple pole at \( s = 1 \). This continuation lies at the heart of analytic number theory: it gives access to the zeros of \( \zeta(s) \), which govern the distribution of primes~\cite{riemann1859ueber}.

The zeros of the Riemann zeta function fall into two classes: the trivial zeros, located at the negative even integers, and the nontrivial zeros, which we denote by \( \rho \) and call \emph{nontrivial Riemann zeros}. Every nontrivial Riemann zero is known to belong to the open critical strip \( 0<\Re(\rho)<1 \)~\cite{edwards1974riemannaes,titchmarsh1986theory}, and the Riemann Hypothesis (RH) asserts that each in fact lies on the critical line:
\[
  \Re(\rho)= 1/2 \, .
\]

Since its first appearance as a brief remark in 1859~\cite{riemann1859ueber}, RH has stood as one of the deepest unsolved problems in mathematics. Among the many equivalent formulations and approaches to RH, the Hilbert--Pólya Conjecture (HPC) stands out for its spectral character: it postulates the existence of a self-adjoint operator with point spectrum
\[
  \sigma_{\mathrm{hp}}:=\left\{i\left(\rho-1/2\right)\mid\zeta(\rho)=0,\;0<\Re(\rho)<1\right\} \, ,
\]
so that the reality of the spectrum, guaranteed by self-adjointness, would force \( \Re(\rho)=1/2 \). An HPC-based approach therefore requires overcoming two principal challenges:
\begin{enumerate}[label=(\Roman*)]
\item constructing an operator whose point spectrum contains \( \sigma_{\mathrm{hp}} \); and
\item proving that this operator, on the specified domain, is self-adjoint.
\end{enumerate}
Several spectral approaches have made partial progress toward these goals~\cite{okubo1998lorentz,connes1999trace,kuipers2014quantum}, notably the Berry--Keating program~\cite{berry1999,bender2017hamiltonian,yakaboylu2024hamiltonian}. A rigorous construction satisfying both requirements, however, remains to be established.

In the present work, building on Ref.~\cite{yakaboylu2024hamiltonian}, we introduce a densely defined non-symmetric operator
\[
  \hat{\mathcal{R}}\colon\mathcal{D}(\hat{\mathcal{R}})\subset L^2([0,\infty),dx)\to L^2([0,\infty),dx) \, ,
\]
whose point spectrum is exactly the Hilbert--Pólya spectral set, \( \sigma_{\mathrm{p}}(\hat{\mathcal{R}})=\sigma_{\mathrm{hp}} \). We refer to \( \hat{\mathcal{R}} \) as the \emph{Riemann operator}. Its eigenfunctions are square-integrable and satisfy biorthogonality relations with corresponding eigenfunctions of the Hilbert-space adjoint \( \hat{\mathcal{R}}^* \), from which we derive our main theorem.

\begin{result}
\label{thm:main}
There exists a linear operator \( \hat{W} \) that intertwines the Riemann operator \( \hat{\mathcal{R}} \) with its Hilbert-space adjoint \( \hat{\mathcal{R}}^* \), such that \( \hat{\mathcal{R}}\,\hat{W}=\hat{W}\,\hat{\mathcal{R}}^* \). Strict positivity of \( \hat{W} \) is equivalent to the Riemann Hypothesis together with the simplicity of the nontrivial zeros:
\[
\hat{W}>0\quad\Longleftrightarrow\quad\Re(\rho)=1/2\;\text{ and }\;\zeta^{(1)}(\rho)\neq0\qquad\text{for every nontrivial zero }\rho \, .
\]
\end{result}

Moreover, the condition \( \hat{W}>0 \) is an operator-theoretic counterpart of Bombieri's refinement of Weil's positivity criterion~\cite{weil1952formules,Bombieri2000}. The same condition is also a prerequisite for a possible construction of a self-adjoint Hilbert--Pólya operator from \( \hat{\mathcal{R}} \) through a similarity transform based on \( \hat{W} \). This recasts challenge~(II) and reverses the usual order of inference: the positivity needed to establish self-adjointness---and thereby deduce RH---already entails RH.

Finally, we discuss possible extensions of the framework to other \( L \)-functions. The existence of a Mellin representation with a smooth, nonvanishing weight and appropriate endpoint behavior provides the basis for constructing the corresponding Riemann-type operator. The intertwiner \( \hat{W} \) additionally requires invariance of the set of nontrivial zeros under reflection about the critical line, a symmetry ensured by a conjugate functional equation.

\section{Preliminaries}
\label{sec_prelim}

We begin by introducing the \emph{completed zeta function}, then outline the Hilbert-space framework and the canonical operators. Throughout the paper, the superscript \( (m) \) applied to a function denotes its \( m \)-th derivative.

\subsection{The Completed Zeta Function}
\label{subsec_completed_zeta}

For \( t > 0 \), define the weight function
\begin{equation}
\label{weight_function}
  \omega(t) := -\frac{d}{dt}\left[t^2 \, \mathrm{Li}_{-1}\!\left(e^{-t}\right)\right] = \frac{t}{4\sinh^2(t/2)}\left[t \coth(t/2) - 2\right] \, ,
\end{equation}
where \( \mathrm{Li}_{-1}(z) = \sum_{n\ge1} n \, z^n \), for \( |z|<1 \), denotes the polylogarithm of order \( -1 \). The closed form follows from \( \mathrm{Li}_{-1}(e^{-t}) = 1/[4\sinh^2(t/2)] \). Since \( t \coth(t/2) > 2 \), the weight is positive on \( (0,\infty) \), and its endpoint behavior is
\begin{subequations}
\label{weight_asymptotics}
\begin{align}
\label{weight_origin}
  \omega(t) &= \frac{t}{6}\left[1 + O(t^2)\right] \quad (t \to 0^+) \, , \\
\label{weight_infinity}
  \omega(t) &= t^2 e^{-t}\left[1 - \frac{2}{t} + O\!\left(e^{-t}\right)\right] \quad (t \to \infty) \, .
\end{align}
\end{subequations}
These estimates follow from the expansions of \( \sinh \) and \( \coth \) at the origin, and from \( 4\sinh^2(t/2) = e^t(1-e^{-t})^2 \) together with \( \coth(t/2) = 1 + 2e^{-t} + O(e^{-2t}) \) at infinity. Its Mellin transform is
\begin{equation}
\label{funct_Z}
  \mathfrak{z}(s) := \int_0^\infty t^{s-1}\,\omega(t)\,dt = \Gamma(s+1)\,(s-1)\,\zeta(s), \qquad \Re(s)>-1 \, .
\end{equation}
The linear vanishing at the origin and the exponential decay at infinity in~\eqref{weight_asymptotics} imply that the integral converges absolutely and defines a holomorphic function on \( \Re(s)>-1 \). Integration by parts and termwise integration establish the identity for \( \Re(s)>1 \). The identity theorem then extends it to \( \Re(s)>-1 \), with the right-hand side understood by continuity at \( s=1 \). The same expression provides a meromorphic continuation of \( \mathfrak{z}(s) \) to \( \mathbb{C} \). We call \( \mathfrak{z}(s) \) the \emph{completed} zeta function, the qualifier referring to the included gamma function \( \Gamma(s+1) \) and the factor \( (s-1) \), which differ from those of the classical completion \( \xi(s) = s(s-1)\,\pi^{-s/2}\,\Gamma(s/2)\,\zeta(s)/2 \).

\begin{definition}[Zero Set of \( \mathfrak{z}(s) \)]
\label{def_zeros}
The gamma function has no zeros, and the simple poles of \( \Gamma(s+1) \) at \( s=-2n \) cancel the trivial zeros of \( \zeta(s) \), giving
\begin{equation}
  \lim_{s \to -2n} \mathfrak{z}(s) = \frac{(1+2n)\,\zeta^{(1)}(-2n)}{(2n-1)!} \neq 0, \qquad n \in \mathbb{Z}^+ \, .
\end{equation}
The remaining poles of \( \Gamma(s+1) \), at \( s=-1,-3,-5,\ldots \), give simple poles of \( \mathfrak{z}(s) \). Furthermore, the zero of the factor \( (s-1) \) is cancelled by the pole of \( \zeta(s) \) at \( s=1 \), leaving
\begin{equation}
  \mathfrak{z}(1) = \Gamma(2)\lim_{s\to1}(s-1)\,\zeta(s) = 1 \, .
\end{equation}
The zeros of the completed zeta function are therefore exactly the nontrivial Riemann zeros, with the same multiplicities. We denote this zero set by
\begin{equation}
\label{rem_zeros}
  \mathcal{Z}_\zeta := \left\{\rho \in \mathbb{C} \mid \zeta(\rho)=0,\; 0<\Re(\rho)<1\right\} \, ,
\end{equation}
where we write \( \rho \) without additional indexing. This set is invariant under the involution \( \rho \mapsto 1-\bar{\rho} \), the composition of complex conjugation \( s \mapsto \bar{s} \) with the reflection \( s \mapsto 1-s \). The former symmetry follows from \( \zeta(\bar{s}) = \overline{\zeta(s)} \), and the latter from the functional equation
\begin{equation}
\label{func_eq}
  \pi^{-s/2}\,\Gamma\left(\frac{s}{2}\right)\,\zeta(s) = \pi^{-(1-s)/2}\,\Gamma\left(\frac{1-s}{2}\right)\,\zeta(1-s) \, ,
\end{equation}
understood as an identity of meromorphic functions.
\end{definition}

\subsection{Hilbert-Space Framework and Canonical Operators}
\label{subsec_framework}

Let
\begin{subequations}
\begin{equation}
  \mathcal{H} := L^2([0,\infty),dx)
\end{equation}
be the Hilbert space of square-integrable functions on the half-line \( [0,\infty) \), and denote its dense subspace of smooth functions compactly supported in \( (0,\infty) \) by
\begin{equation}
  \mathcal{C} := C_c^\infty((0,\infty)) \subset \mathcal{H} \, .
\end{equation}
\end{subequations}
The usual test-function topology on \( \mathcal{C} \) yields the rigged Hilbert space \( \mathcal{C}\subset\mathcal{H}\subset\mathcal{C}' \), where \( \mathcal{C}' \) is the continuous anti-dual of \( \mathcal{C} \).

Consider the multiplication and differentiation operators \( \hat{x},\hat{p} \colon \mathcal{C} \to \mathcal{H} \), defined for \( \psi \in \mathcal{C} \) by \( (\hat{x}\,\psi)(x) = x\,\psi(x) \) and \( (\hat{p}\,\psi)(x) = -i\,\psi^{(1)}(x) \). Both operators preserve \( \mathcal{C} \). As multiplication by \( x \), the operator \( \hat{x} \) is essentially self-adjoint on \( \mathcal{C} \). Its generalized eigenvectors \( \ket{x} \in \mathcal{C}' \) formally satisfy \( \hat{x}\ket{x} = x\ket{x} \), made precise by \( \braket{x|\hat{x}\,\psi} = x\,\braket{x|\psi} \) for all \( \psi \in \mathcal{C} \). The family \( \{\ket{x}\}_{x>0} \) forms a generalized eigenbasis in this rigged Hilbert space. Within this notation, we identify \( \psi \) with \( \ket{\psi} \) via \( \braket{x|\psi} := \psi(x) \), treating \( \bra{x} \) as the evaluation functional on \( \mathcal{C} \) and \( \ket{x} \) as the corresponding anti-linear functional. Whenever an operator \( \hat{q} \) and its formal adjoint \( \hat{q}^\dagger \) map \( \mathcal{C} \) continuously into itself, their dual actions satisfy \( \braket{x|\hat{q}|\psi} = \braket{x|\hat{q}\,\psi} = \braket{\hat{q}^\dagger x|\psi} \). Throughout, \( \hat{q}^\dagger \) denotes the formal adjoint and \( \hat{q}^* \) the Hilbert-space adjoint.

In contrast, \( \hat{p} \) is symmetric on \( \mathcal{C} \) but, with deficiency indices \( (1,0) \), admits no self-adjoint extension in \( \mathcal{H} \).

\begin{definition}[Canonical Operators]
On \( \mathcal{C} \), define the symmetric operators
\begin{equation}
\label{D_T_operators}
  \hat{D} := \frac{1}{2}\left(\hat{x}\,\hat{p} + \hat{p}\,\hat{x}\right) \quad \text{and} \quad \hat{T} := \frac{1}{2}\left(\hat{x}\,\hat{p}^2 + \hat{p}^2\,\hat{x}\right) \, .
\end{equation}
Here, \( \hat{D} \) is the dilation operator, also known as the Berry--Keating Hamiltonian~\cite{berry1999}, whereas we call \( \hat{T} \) the Bessel operator, since in position space its eigenvalue equation is a Bessel-type differential equation; see Proposition~\ref{prop_T}.
\end{definition}

\begin{proposition}[Operator \( \hat{D} \)]
\label{prop_D}
The operator \( \hat{D} \) is essentially self-adjoint on \( \mathcal{C} \), and hence admits a unique self-adjoint extension. We use the same symbol for this extension, whose domain is denoted by \( \mathcal{D}(\hat{D}) \). Its spectrum is simple and purely absolutely continuous, \( \sigma(\hat{D}) = \mathbb{R} \), with generalized eigenfunctions
\begin{equation}
\label{D_eigenbasis}
  \braket{x|\nu} = \frac{1}{\sqrt{2\pi}}\,x^{-1/2+i\nu}, \qquad \hat{D}\ket{\nu} = \nu\ket{\nu}, \qquad \nu \in \mathbb{R} \, ,
\end{equation}
forming a generalized eigenbasis of the rigged Hilbert space.
\end{proposition}

\begin{proof}
In position space the eigenvalue equation reads
\begin{equation}
\label{D_eig}
  \braket{x|\hat{D}\,\nu} = -i\left(x\,\frac{d}{dx} + \frac{1}{2}\right)\braket{x|\nu} = \nu\,\braket{x|\nu} \, ,
\end{equation}
and is solved by~\eqref{D_eigenbasis}. To establish essential self-adjointness, consider the deficiency equations for the adjoint \( \hat{D}^* \) of the operator initially defined on \( \mathcal{C} \). Solving the corresponding differential equations \( (\hat{D}^*\mp i)\psi=0 \) gives formal solutions proportional to \( x^{-3/2} \) and \( x^{1/2} \). These fail square-integrability at the origin and at infinity, respectively, so both deficiency spaces are trivial, and by von Neumann's theorem \( \hat{D} \) is essentially self-adjoint. The spectral properties follow from the Mellin transform, whose inverse has kernel~\eqref{D_eigenbasis}: it is unitary from \( \mathcal{H} \) onto \( L^2(\mathbb{R},d\nu) \) and carries \( \hat{D} \) into multiplication by \( \nu \). The functions~\eqref{D_eigenbasis} satisfy \( \braket{\nu|\nu'} = \delta(\nu-\nu') \) in the distributional sense, so that \( \{\ket{\nu}\}_{\nu\in\mathbb{R}} \subset \mathcal{C}' \) is a generalized eigenbasis. See Ref.~\cite{endres2010berry} for details.
\end{proof}

\begin{proposition}[Operator \( \hat{T} \)]
\label{prop_T}
The operator \( \hat{T} \) is nonnegative on \( \mathcal{C} \) and hence admits the Friedrichs self-adjoint extension, which we again denote by \( \hat{T} \), with domain \( \mathcal{D}(\hat{T}) \). Its spectrum is purely absolutely continuous, \( \sigma(\hat{T}) = [0,\infty) \), with generalized eigenfunctions given by the Bessel function of the first kind of order zero,
\begin{equation}
\label{T_eigenbasis}
  \braket{x|t} = J_0(2\sqrt{xt}), \qquad \hat{T}\ket{t} = t\ket{t}, \qquad t>0 \, ,
\end{equation}
forming a complete generalized eigenbasis.
\end{proposition}

\begin{proof}
In position space, the eigenvalue equation reads
\begin{equation}
\label{T_eig}
  \braket{x|\hat{T}\,t} = \left(-x\,\frac{d^2}{dx^2} - \frac{d}{dx}\right)\braket{x|t} = t\,\braket{x|t} \, .
\end{equation}
At \( t=0 \), the independent solutions are \( 1 \) and \( \log x \). Both are square-integrable near the origin, whereas neither is square-integrable near infinity. The differential expression is therefore in the limit-circle case at \( x=0 \) and the limit-point case at \( x=\infty \). Consequently, the operator initially defined on \( \mathcal{C} \) has deficiency indices \( (1,1) \) and admits a one-parameter family of self-adjoint extensions.

The nonnegativity of \( \hat{T} \) on \( \mathcal{C} \),
\begin{equation}
  \braket{\psi|\hat{T}\,\psi} = \int_0^\infty x\,\bigl|\psi^{(1)}(x)\bigr|^2\,dx \ge 0, \qquad \psi\in\mathcal{C} \, ,
\end{equation}
guarantees the existence of the nonnegative Friedrichs extension~\cite{reed1972methods,teschl2014mathematical}, which we again denote by \( \hat{T} \). Its form domain is the closure of \( \mathcal{C} \) in the associated form norm. In particular, \( \sigma(\hat{T}) \subseteq [0,\infty) \). At the origin, this extension imposes the boundary condition
\begin{equation}
  \lim_{x\to0^+} x\,\psi^{(1)}(x)=0, \qquad \psi\in\mathcal{D}(\hat{T}) \, .
\end{equation}

For \( t>0 \), the substitution \( y=2\sqrt{xt} \) transforms~\eqref{T_eig} into Bessel's differential equation of order zero. The solutions of~\eqref{T_eig} are therefore spanned by \( J_0(2\sqrt{xt}) \) and \( Y_0(2\sqrt{xt}) \). The Friedrichs boundary condition excludes the logarithmic solution \( Y_0 \) and selects the regular solution \( J_0 \), normalized by \( J_0(0)=1 \), as the generalized eigenfunction.

The map \( \psi \mapsto \braket{t|\psi} = \int_0^\infty J_0(2\sqrt{xt})\,\psi(x)\,dx \), initially defined on \( \mathcal{C} \), extends to a unitary Hankel transform of order zero from \( \mathcal{H} \) onto \( L^2((0,\infty),dt) \) and carries the Friedrichs extension \( \hat{T} \) into multiplication by \( t \). Consequently, \( \sigma(\hat{T}) = [0,\infty) \) is purely absolutely continuous. The same Hankel transform gives the distributional completeness relation
\begin{equation}
\label{bessel_dirac}
  \delta(x-x') = \int_0^\infty J_0(2\sqrt{xt})\,J_0(2\sqrt{x't})\,dt \, ,
\end{equation}
which establishes \( \{\ket{t}\}_{t>0} \subset \mathcal{C}' \) as a generalized eigenbasis.
\end{proof}

Moreover, the following identities will be used. On \( \mathcal{C} \), the definitions~\eqref{D_T_operators} give
\begin{equation}
\label{DT_comm}
  [\hat{D},\hat{T}] = i\,\hat{T} \, .
\end{equation}
Since \( J_0(2\sqrt{xt}) \) depends on \( x \) and \( t \) only through their product, the generalized eigenfunctions~\eqref{T_eigenbasis} obey the dilation symmetry
\begin{equation}
\label{xt_relation}
  x\,\partial_x J_0(2\sqrt{xt}) = t\,\partial_t J_0(2\sqrt{xt}) \, .
\end{equation}
Finally, the two generalized eigenbases~\eqref{D_eigenbasis} and~\eqref{T_eigenbasis} are related by the overlap
\begin{equation}
\label{nu_t_overlap}
  \braket{t|\nu} = \frac{1}{\sqrt{2\pi}}\,\frac{\Gamma(1/2+i\nu)}{\Gamma(1/2-i\nu)}\,t^{-1/2-i\nu} \, ,
\end{equation}
which follows from the Mellin integral \( \int_0^\infty x^{s-1}J_0(2\sqrt{xt})\,dx=t^{-s}\Gamma(s)/\Gamma(1-s) \), convergent as an improper integral for \( t>0 \) and \( 0<\Re(s)<3/4 \), evaluated at \( s=1/2+i\nu \). At this value of \( s \), the convergence is conditional. The gamma-function ratio is a pure phase, since \( \Gamma(1/2-i\nu) = \overline{\Gamma(1/2+i\nu)} \) for real \( \nu \). Using~\eqref{xt_relation} and integrating by parts, we obtain the action of the dilation operator in the \( \hat{T} \)-representation,
\begin{equation}
\label{D_in_T_rep}
  \braket{t|\hat{D}\,\psi} = i\left(t\,\frac{d}{dt} + \frac{1}{2}\right)\braket{t|\psi}, \qquad \psi\in\mathcal{C} \, .
\end{equation}
The sign is opposite to that of the position-space action~\eqref{D_eig}, since \( \hat{T} \) scales inversely to \( \hat{x} \) under dilations: compare~\eqref{DT_comm} with \( [\hat{D},\hat{x}] = -i\,\hat{x} \) on \( \mathcal{C} \).

\section{Riemann Operator and Its Spectrum}
\label{sec_Riemann_operator}

We now introduce the Riemann operator \( \hat{\mathcal{R}} \) and determine its point spectrum.

\begin{definition}[Riemann Operator]
\label{def_R}
Let \( \hat{D} \) and \( \hat{T} \) be the self-adjoint operators from Propositions~\ref{prop_D} and~\ref{prop_T}. Define
\begin{equation}
\label{R_op}
  \hat{\mathcal{R}} := \hat{D} + i\,\mu(\hat{T}), \qquad \mu(t) := -t\,\frac{d}{dt}\log\omega(t), \qquad t>0 \, ,
\end{equation}
acting on the dense domain
\begin{equation}
  \mathcal{D}(\hat{\mathcal{R}}) := \left\{\psi \in \mathcal{D}(\hat{D}) \cap \mathcal{D}(\hat{T}) \mid \psi(0)=0\right\} \, .
\end{equation}
\end{definition}

The domain is justified as follows. Since \( \omega \) is smooth and positive on \( (0,\infty) \), the function \( \mu \) is well defined and real-valued there. The endpoint expansions underlying~\eqref{weight_asymptotics} give
\begin{equation}
\label{mu_endpoints}
  \mu(0^+)=-1, \qquad \mu(t)=t-2+O(1/t) \quad (t\to\infty) \, .
\end{equation}
We extend \( \mu \) continuously to \( t=0 \) by setting \( \mu(0):=-1 \). The function \( \mu(t)-t \) has limits \( -1 \) at the origin and \( -2 \) at infinity, and hence is bounded and continuous on \( [0,\infty) \). Consequently, the functional calculus of \( \hat{T} \) decomposes \( \mu(\hat{T})=\hat{T}+(\mu(\hat{T})-\hat{T}) \) into an unbounded part with domain \( \mathcal{D}(\hat{T}) \) and a remainder that extends to a bounded self-adjoint operator on \( \mathcal{H} \). This bounded perturbation preserves the domain, so
\begin{equation}
\mathcal{D}(\mu(\hat{T}))=\mathcal{D}(\hat{T}) \, .
\end{equation}
Thus, before imposing the boundary condition, the sum \( \hat{D}+i\,\mu(\hat{T}) \) is naturally defined on \( \mathcal{D}(\hat{D})\cap\mathcal{D}(\hat{T}) \).

For the boundary condition, note that essential self-adjointness of the dilation operator \( \hat{D} \) on \( \mathcal{C} \) requires no additional condition at the origin, and a finite limit there need not exist for every \( \psi\in\mathcal{D}(\hat{D}) \). The Friedrichs realization of \( \hat{T} \), however, ensures a finite boundary trace throughout \( \mathcal{D}(\hat{T}) \). Indeed, let \( \psi\in\mathcal{D}(\hat{T}) \) and set \( g:=\hat{T}\psi\in\mathcal{H} \). The differential equation \( g=-(x\psi^{(1)})^{(1)} \), together with the Friedrichs condition \( \lim_{x\to0^+}x\psi^{(1)}(x)=0 \), gives
\begin{equation}
  x\,\psi^{(1)}(x)=-\int_0^x g(u)\,du, \qquad \bigl|\psi^{(1)}(x)\bigr| \le x^{-1/2}\,\|g\|_{L^2(0,x)} \, .
\end{equation}
Consequently, \( \psi^{(1)} \) is integrable near the origin, so the boundary value \( \psi(0):=\lim_{x\to0^+}\psi(x) \) exists as a finite limit.

This boundary value is not fixed by the Friedrichs condition. For example, \( \psi(x)=e^{-x} \) belongs to \( \mathcal{D}(\hat{D})\cap\mathcal{D}(\hat{T}) \) and has \( \psi(0)=1 \). The additional Dirichlet condition in the definition of \( \hat{\mathcal{R}} \) therefore selects a proper subspace of this intersection. Since \( \mathcal{C}\subset\mathcal{D}(\hat{D})\cap\mathcal{D}(\hat{T}) \) and every \( \psi\in\mathcal{C} \) vanishes near the origin, we have \( \mathcal{C}\subset\mathcal{D}(\hat{\mathcal{R}}) \); in particular, \( \hat{\mathcal{R}} \) is densely defined.

\begin{lemma}[Formal Intertwiner]
\label{lemma_similarity}
Formally, the commutation relation~\eqref{DT_comm} and the definition of \( \mu \) in~\eqref{R_op} yield the factorization
\begin{equation}
\label{R_factorized}
  \hat{\mathcal{R}} = \omega(\hat{T})\,\hat{D}\,\omega(\hat{T})^{-1} \, ,
\end{equation}
and hence \( \hat{\mathcal{R}}^\dagger = \omega(\hat{T})^{-1}\,\hat{D}\,\omega(\hat{T}) \). Consequently, for every real-valued Borel function \( \mathcal{V} \) on \( \sigma(\hat{D})=\mathbb{R} \), the formally symmetric expression
\begin{equation}
\label{V_general}
  \hat{V} := \omega(\hat{T})\,\mathcal{V}(\hat{D})\,\omega(\hat{T})
\end{equation}
satisfies the formal intertwining identity
\begin{equation}
\label{V_intertwining}
  \hat{\mathcal{R}}\,\hat{V} = \hat{V}\,\hat{\mathcal{R}}^\dagger \, .
\end{equation}
Note that these formal identities imply neither equality of operator domains nor compatibility with the Dirichlet condition in Definition~\ref{def_R}.
\end{lemma}

\begin{proof}
The weight function \( \omega \) is bounded, being continuous and tending to zero at both endpoints, so functional calculus defines a bounded self-adjoint operator \( \omega(\hat{T}) \). Since \( \omega(t)>0 \) for \( t>0 \) and \( 0 \) is not an eigenvalue of \( \hat{T} \), this operator is injective and has dense range. Its inverse is therefore densely defined on \( \operatorname{Ran}\omega(\hat{T}) \), although this domain need not contain \( \mathcal{C} \). In the \( \hat{T} \)-representation, the inverse acts as multiplication by \( \omega(t)^{-1} \) on its domain.

Working formally, the conjugation of \( \hat{D} \) by \( \omega(\hat{T}) \) can be written as
\begin{equation}
  \omega(\hat{T})\,\hat{D}\,\omega(\hat{T})^{-1} = \hat{D} - [\hat{D},\omega(\hat{T})]\,\omega(\hat{T})^{-1} \, .
\end{equation}
Using the differential expression~\eqref{D_in_T_rep} in the \( \hat{T} \)-representation, we obtain
\begin{equation}
\label{D_omega_comm}
  [\hat{D},\omega(\hat{T})] = i\,\hat{T}\,\omega^{(1)}(\hat{T}) \, .
\end{equation}
Thus \( -[\hat{D},\omega(\hat{T})]\,\omega(\hat{T})^{-1} = i\,\mu(\hat{T}) \) by~\eqref{R_op}, establishing the formal factorization~\eqref{R_factorized}. Taking formal adjoints gives \( \hat{\mathcal{R}}^\dagger = \hat{D}-i\,\mu(\hat{T}) = \omega(\hat{T})^{-1}\,\hat{D}\,\omega(\hat{T}) \).

Since \( \mathcal{V} \) is real-valued, \( \mathcal{V}(\hat{D}) \) is self-adjoint, and the expression~\eqref{V_general} is formally symmetric. The formal intertwining identity follows from the spectral commutation of \( \hat{D} \) and \( \mathcal{V}(\hat{D}) \):
\begin{equation}
  \hat{\mathcal{R}}\,\hat{V} = \omega(\hat{T})\,\hat{D}\,\mathcal{V}(\hat{D})\,\omega(\hat{T}) = \omega(\hat{T})\,\mathcal{V}(\hat{D})\,\hat{D}\,\omega(\hat{T}) = \hat{V}\,\hat{\mathcal{R}}^\dagger \, .
\end{equation}
\end{proof}

\begin{theorem}[Eigensystem of \( \hat{\mathcal{R}} \)]
\label{theo_eigen_R}
The point spectrum of \( \hat{\mathcal{R}} \) is
\begin{equation}
\label{spec_R}
  \sigma_{\mathrm{p}}(\hat{\mathcal{R}}) = \left\{i\left(\rho-1/2\right) \mid \rho\in\mathcal{Z}_\zeta\right\} \, ,
\end{equation}
with the associated eigenstates
\begin{equation}
\label{eigen_R}
  \ket{\Psi_\rho} = \int_0^\infty t^{\rho-1}\,\omega(t)\ket{t}\,dt \, .
\end{equation}
\end{theorem}

\begin{proof}
We write the eigenvalue equation for \( \hat{\mathcal{R}} \) in the form
\begin{equation}
\label{eigen_ket}
  \hat{\mathcal{R}}\,\ket{\Psi_s}=i\left(s-1/2\right)\ket{\Psi_s} \, .
\end{equation}
Expanding in the generalized eigenbasis \( \{\ket{t}\}_{t>0} \) of \( \hat{T} \), we write
\begin{equation}
  \ket{\Psi_s}=\int_0^\infty f_s(t)\ket{t}\,dt, \qquad f_s(t):=\braket{t|\Psi_s} \, .
\end{equation}
Under the unitary Hankel transform, formula~\eqref{D_in_T_rep} extends by closure to the self-adjoint domain of \( \hat{D} \). Thus any eigenstate in \( \mathcal{D}(\hat{\mathcal{R}}) \) has a locally absolutely continuous coefficient \( f_s \) satisfying \( f_s,tf_s,tf_s^{(1)}\in L^2((0,\infty),dt) \). The Cauchy--Schwarz inequality then gives \( f_s\in L^1((0,\infty),dt) \), the space of absolutely integrable functions.

The corresponding position-space representation is
\begin{equation}
  \Psi_s(x):=\braket{x|\Psi_s}=\int_0^\infty J_0(2\sqrt{xt})\,f_s(t)\,dt \, .
\end{equation}
This integral is absolutely convergent since \( |J_0(2\sqrt{xt})|\le1 \). Using~\eqref{xt_relation}, integration by parts gives
\begin{equation}
\label{int_eq}
  \int_0^\infty J_0(2\sqrt{xt})\left[it\frac{d}{dt}+\frac{i}{2}+i\mu(t)-i\left(s-\frac{1}{2}\right)\right]f_s(t)\,dt+\mathcal{B}[f_s]=0 \, ,
\end{equation}
where the integral is understood as an \( L^2 \) Hankel transform and the boundary term is
\begin{equation}
\label{bc_eigen_ket}
  \mathcal{B}[f_s]=-i\left.tf_s(t)J_0(2\sqrt{xt})\right|_{t=0}^{\infty} \, .
\end{equation}

The product \( tf_s \) also belongs to the Sobolev space \( H^1((0,\infty)) \), consisting of square-integrable functions whose first weak derivative is square-integrable, since \( (tf_s)^{(1)}=f_s+tf_s^{(1)}\in L^2((0,\infty)) \). It therefore tends to zero at infinity and has a finite limit at the origin. The latter limit must also be zero because \( f_s \) is square-integrable. Hence the boundary term~\eqref{bc_eigen_ket} vanishes.

The differential expression
\begin{equation}
\label{diff_t_space}
  \mathcal{L}_s f_s := \left[it\frac{d}{dt}+\frac{i}{2}+i\mu(t)-i\left(s-\frac{1}{2}\right)\right]f_s
\end{equation}
belongs to \( L^2((0,\infty),dt) \) by the domain conditions and the decomposition \( \mu f_s=tf_s+(\mu-t)f_s \), since \( \mu(t)-t \) is bounded. By~\eqref{int_eq} and injectivity of the unitary Hankel transform, \( \mathcal{L}_s f_s=0 \) almost everywhere. Using \( \mu(t)=-t\omega^{(1)}(t)/\omega(t) \), this first-order ODE has the solution
\begin{equation}
\label{F_sol}
  f_s(t)=t^{s-1}\,\omega(t) \, ,
\end{equation}
unique up to a nonzero overall constant, which we set to one. Since \( f_s(t)\sim t^s/6 \) at the origin and decays exponentially at infinity, this solution is square-integrable precisely when \( \Re(s)>-1/2 \). Every eigenstate must therefore satisfy this condition.

For \( \Re(s)>-1/2 \), the position-space representation is
\begin{equation}
\label{eigen_func}
\Psi_s(x)=\int_0^\infty J_0(2\sqrt{xt})\,t^{s-1}\,\omega(t)\,dt \, .
\end{equation}
This integral converges absolutely: the integrand behaves as \( t^s/6 \) at the origin and decays exponentially at infinity. Moreover, \( |J_0(2\sqrt{xt})|\le1 \), and for each fixed \( t>0 \), \( J_0(2\sqrt{xt})\to0 \) as \( x\to\infty \). Dominated convergence therefore gives \( \Psi_s(x)\to0 \).

Imposing the Dirichlet condition at the origin gives
\begin{equation}
\label{int_bc}
\Psi_s(0)=\int_0^\infty t^{s-1}\,\omega(t)\,dt=\mathfrak{z}(s)=0 \, ,
\end{equation}
where \( J_0(0)=1 \) was used. By~\eqref{rem_zeros}, any eigenstate must therefore have \( s=\rho\in\mathcal{Z}_\zeta \). It remains to verify that every such zero produces a nonzero eigenstate in \( \mathcal{D}(\hat{\mathcal{R}}) \). Since \( \Re(\rho)>0 \) for all \( \rho\in\mathcal{Z}_\zeta \), the coefficient \( f_\rho(t)=t^{\rho-1}\omega(t) \) is square-integrable, and unitarity of the Hankel transform gives
\begin{equation}
  \int_0^\infty|\Psi_\rho(x)|^2\,dx=\int_0^\infty|f_\rho(t)|^2\,dt=\int_0^\infty t^{2\Re(\rho)-2}\,\omega(t)^2\,dt<\infty \, .
\end{equation}
In the \( \hat{T} \)-representation, \( \hat{T} \) multiplies \( f_\rho \) by \( t \), while~\eqref{diff_t_space} with \( \mathcal{L}_\rho f_\rho=0 \) gives
\begin{equation}
  i\left(t\frac{d}{dt}+\frac{1}{2}\right)f_\rho(t)=i\left[\rho-\frac{1}{2}-\mu(t)\right]f_\rho(t) \, .
\end{equation}
Both \( tf_\rho \) and the right-hand side are square-integrable: their multipliers are bounded near the origin and grow at most linearly at infinity, where \( f_\rho \) decays exponentially. Hence \( \ket{\Psi_\rho}\in\mathcal{D}(\hat{D})\cap\mathcal{D}(\hat{T}) \). Together with \( \Psi_\rho(0)=\mathfrak{z}(\rho)=0 \), this proves \( \ket{\Psi_\rho}\in\mathcal{D}(\hat{\mathcal{R}}) \) and establishes the point spectrum~\eqref{spec_R} with the eigenstates~\eqref{eigen_R}.
\end{proof}

\begin{remark}
\label{gen_R_op}
The construction of the Riemann operator \( \hat{\mathcal{R}} \) is not unique. By the argument of Theorem~\ref{theo_eigen_R}, one may replace \( \omega(t) \) in~\eqref{weight_function} with a positive weight \( \omega'(t) \) whose Mellin transform vanishes on \( \mathcal{Z}_\zeta \), provided the resulting operator is well defined and the corresponding eigenfunctions belong to its domain. The formal factorization~\eqref{R_factorized} then yields the modified Riemann operator \( \hat{\mathcal{R}}':=\hat{D}+i\,\mu'(\hat{T}) \), with \( \mu' \) obtained from \( \omega' \) via~\eqref{R_op}.

To begin with, the weight~\eqref{weight_function} is itself the case \( \alpha=1 \) of a one-parameter family with \( 1/2\le\alpha<2 \),
\begin{equation}
\label{omega_alpha}
  \omega_\alpha(t) := -\frac{d}{dt}\left[t^{\alpha+1}\,\mathrm{Li}_{-\alpha}\!\left(e^{-t}\right)\right] \, .
\end{equation}
These weights inherit positivity on \( (0,\infty) \) from the auxiliary weight at \( \alpha=2 \) through a fractional-integral representation with a positive kernel. Their Mellin transforms are
\begin{equation}
  \mathfrak{z}_\alpha(s) := \int_0^\infty t^{s-1}\,\omega_\alpha(t)\,dt = \Gamma(s+\alpha)\,(s-1)\,\zeta(s), \qquad \Re(s)>-\alpha \, .
\end{equation}
Within this half-plane of convergence, the zero set of \( \mathfrak{z}_\alpha \) is \( \mathcal{Z}_\zeta \): the trivial zeros of \( \zeta(s) \) lie outside it, and the zero of \( (s-1) \) is again cancelled by the pole at \( s=1 \), where \( \mathfrak{z}_\alpha(1)=\Gamma(1+\alpha)\neq0 \). Since \( \omega_\alpha(t)\sim-(\alpha+1)\,\zeta(-\alpha)\,t^\alpha \) at the origin, with exponential decay at infinity, the corresponding eigenfunctions are square-integrable precisely when \( \Re(s)>1/2-\alpha \). For the stated range \( 1/2\le\alpha<2 \), this condition holds throughout the critical strip \( 0<\Re(s)<1 \). Moreover, the logarithmic derivative \( \mu_\alpha(t):=-t\,d\log\omega_\alpha(t)/dt \) satisfies
\begin{equation}
  \mu_\alpha(0^+)=-\alpha, \qquad \mu_\alpha(t)=t-(\alpha+1)+O(1/t) \quad (t\to\infty) \, .
\end{equation}
Thus \( \mu_\alpha(t)-t \) is bounded, and \( \mathcal{D}(\mu_\alpha(\hat{T}))=\mathcal{D}(\hat{T}) \). Hence Theorem~\ref{theo_eigen_R} carries over with the same operator domain and point spectrum.

Beyond this family, the zero set itself may be enlarged. For the same parameter range \( \alpha\in[1/2,2) \), the weight
\begin{equation}
\label{omega_eta}
\tilde{\omega}_\alpha(t) := -t^\alpha\,\mathrm{Li}_{-\alpha}\!\left(-e^{-t}\right) = t^\alpha\sum_{n=1}^\infty(-1)^{n-1}n^\alpha e^{-nt}
\end{equation}
is positive by the same fractional-integral argument and has Mellin transform
\begin{equation}
\label{funct_H}
  \mathfrak{h}_\alpha(s) := \int_0^\infty t^{s-1}\,\tilde{\omega}_\alpha(t)\,dt = \Gamma(s+\alpha)\left(1-2^{1-s}\right)\zeta(s), \qquad \Re(s)>-\alpha \, .
\end{equation}
Within this half-plane of convergence, its zero set is \( \mathcal{Z}_\zeta\cup\mathcal{Z}_{\mathrm{per}} \), where
\begin{equation}
\label{per_zeros}
  \mathcal{Z}_{\mathrm{per}} := \left\{1+\frac{2\pi i k}{\log 2}\;\middle|\;k\in\mathbb{Z}\setminus\{0\}\right\}
\end{equation}
collects the zeros of \( (1-2^{1-s}) \) on the line \( \Re(s)=1 \). The value \( k=0 \) is excluded because cancellation with the pole of \( \zeta(s) \) leaves \( \mathfrak{h}_\alpha(1)=\Gamma(1+\alpha)\log 2\neq0 \). Since \( \tilde{\omega}_\alpha(t)\sim(1-2^{1+\alpha})\zeta(-\alpha)\,t^\alpha \) at the origin, the square-integrability threshold is again \( \Re(s)>1/2-\alpha \). Its logarithmic derivative \( \tilde{\mu}_\alpha(t):=-t\,d\log\tilde{\omega}_\alpha(t)/dt \) satisfies
\begin{equation}
  \tilde{\mu}_\alpha(0^+)=-\alpha, \qquad \tilde{\mu}_\alpha(t)=t-\alpha+O(te^{-t}) \quad (t\to\infty) \, .
\end{equation}
Hence the functional-calculus domain again remains \( \mathcal{D}(\hat{T}) \). The eigenfunctions corresponding to \( \mathcal{Z}_{\mathrm{per}} \) also belong to the operator domain, so the point spectrum is enlarged accordingly.

Moreover, the Bessel operator \( \hat{T} \) may be replaced by another self-adjoint operator whose generalized eigenbasis obeys the dilation symmetry~\eqref{xt_relation}, possibly with a constant scaling factor, and has a nonzero boundary value at \( x=0 \) independent of the spectral parameter. With an appropriate adjustment of the profile and logarithmic-derivative term, the Dirichlet condition can again reduce to \( \mathfrak{z}(s)=0 \), provided the spectral measure and operator domains are compatible with the construction.

A natural instance is \( \hat{T}\to\hat{p}^2 \), where \( \hat{p}^2 \) denotes the Neumann self-adjoint realization of \( -d^2/dx^2 \), with cosine basis \( \braket{x|p}=\sqrt{2/\pi}\cos(px) \), \( p>0 \). Rewriting the boundary Mellin integral under \( t=p^2 \), with \( dt=2p\,dp \), gives the profile \( \braket{p|\Psi_s^{\mathrm{cos}}}=2p^{2s-1}\omega(p^2) \), so that \( \Psi_s^{\mathrm{cos}}(0)=\sqrt{2/\pi}\,\mathfrak{z}(s) \). In this representation, \( \hat{D} \) acts as \( i(p\,d/dp+1/2) \), so the modified operator \( \hat{\mathcal{R}}_{\mathrm{cos}}:=\hat{D}+i[2\mu(\hat{p}^2)-1/2] \) has eigenvalue \( 2i(s-1/2) \) on this profile. Its domain is \( \{\psi\in\mathcal{D}(\hat{D})\cap\mathcal{D}(\hat{p}^2)\mid\psi(0)=0\} \). The profiles are square-integrable precisely when \( \Re(s)>-3/4 \), and for every \( s=\rho\in\mathcal{Z}_\zeta \) they satisfy the operator-domain requirements. Thus the quantizing zero set remains \( \mathcal{Z}_\zeta \), while the point spectrum becomes \( \{2i(\rho-1/2)\mid\rho\in\mathcal{Z}_\zeta\} \).
\end{remark}

\subsection{Adjoint Eigensystem and Biorthogonality}
\label{subsec_adjoint}

Since \( \hat{\mathcal{R}} \) is non-symmetric, its eigenstates \( \{\ket{\Psi_\rho}\}_{\rho\in\mathcal{Z}_\zeta} \) need not be mutually orthogonal. They do, however, satisfy biorthogonality relations with corresponding eigenstates of the Hilbert-space adjoint \( \hat{\mathcal{R}}^* \), which we construct in this subsection.

\begin{theorem}[Dual Eigensystem]
\label{dual_eigensystem}
For every \( \rho\in\mathcal{Z}_\zeta \), the conjugate eigenvalue \( -i(\bar{\rho}-1/2) \) belongs to the point spectrum of \( \hat{\mathcal{R}}^* \), so that \( \overline{\sigma_{\mathrm{p}}(\hat{\mathcal{R}})}\subseteq\sigma_{\mathrm{p}}(\hat{\mathcal{R}}^*) \), with corresponding eigenstates
\begin{equation}
\label{eigen_R_adj}
  \ket{\Phi_\rho} = \kappa_\rho\int_0^\infty \frac{t^{-\bar{\rho}}}{\omega(t)}\left(\int_t^\infty \tau^{\bar{\rho}-1}\,\omega(\tau)\,d\tau\right)\ket{t}\,dt, \qquad \kappa_\rho\in\mathbb{C}^\times \, .
\end{equation}
\end{theorem}

\begin{proof}
By the definition of the Hilbert-space adjoint, it suffices to construct a nonzero \( \ket{\Phi_\rho}\in\mathcal{H} \) satisfying
\begin{equation}
\label{adj_target}
  \braket{\Phi_\rho|\hat{\mathcal{R}}\,U} = i\left(\rho-1/2\right)\braket{\Phi_\rho|U} \qquad \text{for every } \ket{U}\in\mathcal{D}(\hat{\mathcal{R}}) \, .
\end{equation}
This identity establishes both \( \ket{\Phi_\rho}\in\mathcal{D}(\hat{\mathcal{R}}^*) \) and \( \hat{\mathcal{R}}^*\ket{\Phi_\rho}=-i(\bar{\rho}-1/2)\ket{\Phi_\rho} \).

Write \( g_\rho(t):=\braket{t|\Phi_\rho} \) and \( u(t):=\braket{t|U} \). For a locally absolutely continuous candidate \( g_\rho \), integration by parts on finite intervals shows that it suffices to verify
\begin{equation}
\label{adj_target_t}
  \int_0^\infty \overline{\mathcal{L}_\rho^\dagger g_\rho(t)}\,u(t)\,dt+i\left.t\,\overline{g_\rho(t)}\,u(t)\right|_0^\infty=0 \, ,
\end{equation}
provided the integrals and endpoint limits exist, as will be checked below. Here \( \mathcal{L}_\rho^\dagger \) is the formal adjoint of \( \mathcal{L}_s \) in~\eqref{diff_t_space} at \( s=\rho \).

The homogeneous equation \( \mathcal{L}_\rho^\dagger g_\rho=0 \) has the solution \( g_\rho(t)\propto t^{-\bar{\rho}}/\omega(t) \), which grows like \( t^{-\bar{\rho}-2}e^t \) at infinity and hence does not belong to \( \mathcal{H} \). We therefore seek a solution for which the integral in~\eqref{adj_target_t} vanishes through the boundary condition on \( U \). Since \( U\in\mathcal{D}(\hat{T}) \), both \( u \) and \( tu \) are square-integrable, so \( u \) is integrable by the Cauchy--Schwarz inequality. Evaluating the inverse Hankel transform at \( x=0 \) gives \( \int_0^\infty u(t)\,dt=U(0)=0 \). It therefore suffices to take \( \mathcal{L}_\rho^\dagger g_\rho \) to be a nonzero constant:
\begin{equation}
\label{diff_for_G}
  \mathcal{L}_\rho^\dagger g_\rho = \left[it\frac{d}{dt}+\frac{i}{2}-i\mu(t)+i\left(\bar{\rho}-\frac{1}{2}\right)\right]g_\rho = -i\kappa_\rho, \qquad \kappa_\rho\in\mathbb{C}^\times \, .
\end{equation}
Using the integrating factor \( t^{\bar{\rho}}\omega(t) \) and discarding the homogeneous term, which is not square-integrable at infinity, we obtain
\begin{equation}
\label{G_sol}
  g_\rho(t)=\braket{t|\Phi_\rho}=\kappa_\rho\,\frac{t^{-\bar{\rho}}}{\omega(t)}\int_t^\infty \tau^{\bar{\rho}-1}\,\omega(\tau)\,d\tau \, .
\end{equation}
By~\eqref{weight_infinity}, \( g_\rho(t)\sim\kappa_\rho/t \) at infinity. At the origin, the identity \( \mathfrak{z}(\bar{\rho})=\overline{\mathfrak{z}(\rho)}=0 \) gives
\begin{equation}
  \int_t^\infty \tau^{\bar{\rho}-1}\,\omega(\tau)\,d\tau=-\int_0^t \tau^{\bar{\rho}-1}\,\omega(\tau)\,d\tau\sim-\frac{t^{\bar{\rho}+1}}{6(\bar{\rho}+1)} \, .
\end{equation}
Together with~\eqref{weight_origin}, this yields \( g_\rho(t)\to-\kappa_\rho/(\bar{\rho}+1) \). Thus \( g_\rho \) is bounded and square-integrable, and \( \ket{\Phi_\rho} \) is a nonzero element of \( \mathcal{H} \).

Finally, we verify that the boundary term in~\eqref{adj_target_t} vanishes. By the endpoint argument in the proof of Theorem~\ref{theo_eigen_R}, \( tu(t)\to0 \) at both endpoints. Since \( g_\rho \) is bounded, the boundary term vanishes. Passing to the endpoint limits in the finite-interval integration-by-parts identity therefore yields~\eqref{adj_target}, proving the claim.
\end{proof}

\begin{remark}
\label{residual_remark}
The inclusion \( \overline{\sigma_{\mathrm{p}}(\hat{\mathcal{R}})}\subseteq\sigma_{\mathrm{p}}(\hat{\mathcal{R}}^*) \) is strict. Replacing \( \rho \) by an arbitrary \( s\in\mathbb{C} \) in~\eqref{G_sol} defines a coefficient \( g_s(t) \) for every \( t>0 \), with \( \kappa_s\in\mathbb{C}^\times \). Since \( g_s(t)\sim\kappa_s/t \) at infinity, its square-integrability is determined by its behavior at the origin. For \( \Re(s)>-1 \) and \( s\notin\mathcal{Z}_\zeta \), we have \( g_s(t)\sim6\kappa_s\mathfrak{z}(\bar{s})\,t^{-\bar{s}-1} \), with \( \mathfrak{z}(\bar{s})\neq0 \). For \( \Re(s)<-1 \), it tends to \( -\kappa_s/(\bar{s}+1) \), while on the line \( \Re(s)=-1 \) it grows at most logarithmically. Thus, for \( s\notin\mathcal{Z}_\zeta \), the coefficient is square-integrable precisely when \( \Re(s)<-1/2 \).

For these parameters, the Dirichlet condition again annihilates the constant source. The boundary term also vanishes: for \( U\in\mathcal{D}(\hat{\mathcal{R}}) \), the domain conditions give \( t\braket{t|U}=O(t^{1/2}) \) at the origin and \( t\braket{t|U}\to0 \) at infinity, which suffice together with the preceding estimates. The adjoint identity therefore holds, yielding \( \{z\in\mathbb{C}\mid\Im(z)>1\}\subseteq\sigma_{\mathrm{p}}(\hat{\mathcal{R}}^*) \) under \( z=-i(\bar{s}-1/2) \). This half-plane is disjoint from \( \overline{\sigma_{\mathrm{p}}(\hat{\mathcal{R}})} \), which lies in \( |\Im(z)|<1/2 \).

For every such \( z \), the operator \( \hat{\mathcal{R}}-\bar{z} \) is injective because \( \bar{z}\notin\sigma_{\mathrm{p}}(\hat{\mathcal{R}}) \), while \( \overline{\operatorname{Ran}(\hat{\mathcal{R}}-\bar{z})}=\ker(\hat{\mathcal{R}}^*-z)^\perp \) shows that its range is non-dense. Hence the residual spectrum of \( \hat{\mathcal{R}} \) contains every complex number with imaginary part less than \( -1 \). We do not pursue this residual spectrum further, since the associated adjoint eigenstates have no corresponding Hilbert-space eigenstates of \( \hat{\mathcal{R}} \): \( \ket{\Psi_s}\in\mathcal{H} \) requires \( \Re(s)>-1/2 \).
\end{remark}

\begin{remark}
\label{green_remark}
With its constant source term, the differential equation~\eqref{diff_for_G} corresponds to the following distributional eigenvalue equation:
\begin{equation}
\label{general_eigenvalue_eq_adjoint}
  \hat{\mathcal{R}}^\dagger\ket{\Phi_\rho} = -i\left(\bar{\rho}-1/2\right)\ket{\Phi_\rho}-i\kappa_\rho\ket{\delta}, \qquad \ket{\delta}:=\int_0^\infty\ket{t}\,dt \, .
\end{equation}
Here \( \ket{\delta} \) denotes the boundary evaluation distribution on \( C_c^\infty([0,\infty)) \), the space of compactly supported test functions smooth up to \( x=0 \). Its action is defined by \( \braket{\varphi|\delta}=\overline{\varphi(0)} \). In position space, \( \braket{x|\delta}=\delta(x) \), where \( \delta(x) \) denotes the unit boundary delta on the half-line. Its restriction to the original test space \( \mathcal{C}=C_c^\infty((0,\infty)) \) is zero. Thus \( \braket{x|\Phi_\rho} \) satisfies the eigenvalue equation for the \emph{formal} adjoint up to a boundary source supported at \( x=0 \). Since every element of \( \mathcal{D}(\hat{\mathcal{R}}) \) vanishes there, this source annihilates the operator domain.

The same boundary source drops out of Green's identity. For every \( \rho'\in\mathcal{Z}_\zeta \), we have \( \braket{\Psi_{\rho'}|\delta}=\overline{\Psi_{\rho'}(0)}=0 \) by~\eqref{int_bc}. Using the eigenvalue equations therefore gives
\begin{equation}
\braket{\Psi_{\rho'}|\hat{\mathcal{R}}^\dagger\,\Phi_\rho}-\braket{\hat{\mathcal{R}}\,\Psi_{\rho'}|\Phi_\rho}=i(\bar{\rho}'-\bar{\rho})\braket{\Psi_{\rho'}|\Phi_\rho} \, ,
\end{equation}
where the first term is interpreted as a distributional pairing. Because the boundary source pairs to zero, that term equals \( \braket{\Psi_{\rho'}|\hat{\mathcal{R}}^*\,\Phi_\rho} \). The left-hand side consequently vanishes by the defining identity of the Hilbert-space adjoint. Thus \( \braket{\Psi_{\rho'}|\Phi_\rho}=0 \) whenever \( \rho'\neq\rho \), a relation we also verify directly in the next lemma.
\end{remark}

\begin{lemma}[Biorthogonality Relation]
\label{lemma_biorthogonality}
For any two zeros \( \rho,\rho'\in\mathcal{Z}_\zeta \), the eigenstates \( \ket{\Psi_\rho} \) and \( \ket{\Phi_{\rho'}} \) satisfy the biorthogonality relation
\begin{equation}
\label{eq_biorthogonality}
  \braket{\Phi_{\rho'}|\Psi_\rho}=N_\rho\,\delta_{\rho'\rho} \, ,
\end{equation}
with coefficient
\begin{equation}
\label{N_def}
  N_\rho:=\overline{\kappa_\rho}\,\mathfrak{z}^{(1)}(\rho)=\overline{\kappa_\rho}\,\Gamma(\rho+1)\,(\rho-1)\,\zeta^{(1)}(\rho) \, ,
\end{equation}
where the last equality follows from \( \zeta(\rho)=0 \).
\end{lemma}

\begin{proof}
By unitarity of the Hankel transform and~\eqref{eigen_R} and~\eqref{eigen_R_adj}, the inner product reduces to
\begin{equation}
\label{overlap_reduced}
  \braket{\Phi_{\rho'}|\Psi_\rho}=\overline{\kappa_{\rho'}}\int_0^\infty t^{\rho-\rho'-1}\left(\int_t^\infty\tau^{\rho'-1}\,\omega(\tau)\,d\tau\right)dt \, .
\end{equation}
This integral is absolutely convergent by square-integrability of the two spectral coefficients. For \( \rho\neq\rho' \), integration by parts gives
\begin{equation}
  \braket{\Phi_{\rho'}|\Psi_\rho}=\frac{\overline{\kappa_{\rho'}}}{\rho-\rho'}\left(\left.t^{\rho-\rho'}\int_t^\infty\tau^{\rho'-1}\,\omega(\tau)\,d\tau\right|_{t=0}^{\infty}+\int_0^\infty t^{\rho-1}\,\omega(t)\,dt\right)=0 \, .
\end{equation}
The boundary term vanishes at both endpoints. As \( t\to\infty \), the asymptotics~\eqref{weight_infinity} give \( \int_t^\infty\tau^{\rho'-1}\omega(\tau)\,d\tau\sim t^{\rho'+1}e^{-t} \), so the boundary expression is asymptotic to \( t^{\rho+1}e^{-t}\to0 \). As \( t\to0^+ \), the identity \( \mathfrak{z}(\rho')=0 \) and~\eqref{weight_origin} give \( \int_t^\infty\tau^{\rho'-1}\omega(\tau)\,d\tau\sim-t^{\rho'+1}/[6(\rho'+1)] \), so the boundary expression again tends to zero. The remaining integral vanishes because \( \mathfrak{z}(\rho)=0 \).

For \( \rho'=\rho \), integration by parts with the antiderivative \( \log t \) instead yields
\begin{equation}
\begin{split}
  \braket{\Phi_\rho|\Psi_\rho} &= \overline{\kappa_\rho}\int_0^\infty t^{\rho-1}\,\omega(t)\log t\,dt \\
  &=\overline{\kappa_\rho}\left.\frac{d}{ds}\int_0^\infty t^{s-1}\,\omega(t)\,dt\right|_{s=\rho}=\overline{\kappa_\rho}\,\mathfrak{z}^{(1)}(\rho)=N_\rho \, .
\end{split}
\end{equation}
Here, the logarithmic boundary term \( \left.\log t\int_t^\infty\tau^{\rho-1}\omega(\tau)\,d\tau\right|_{t=0}^{\infty} \) vanishes by the same endpoint estimates. Differentiation under the integral sign is justified by dominated convergence for \( s \) near \( \rho \). By~\eqref{N_def}, \( N_\rho \) vanishes precisely when \( \rho \) is a multiple zero.
\end{proof}

\begin{remark}
\label{rem_jordan}
The vanishing of \( N_\rho \) at a multiple zero is not accidental: it signals a nontrivial Jordan chain of \( \hat{\mathcal{R}} \) at the corresponding eigenvalue. Namely, a rank-\( 2 \) generalized eigenvector \( \ket{\Psi_\rho^{\mathrm{r}2}} \), defined by
\begin{equation}
  \left(\hat{\mathcal{R}}-i \left(\rho - 1/2\right) \right)\ket{\Psi_\rho^{\mathrm{r}2}} = \ket{\Psi_\rho} \, ,
\end{equation}
corresponds in the eigenbasis \( \{\ket{t}\}_{t>0} \) to \( \mathcal{L}_\rho \braket{t|\Psi_\rho^{\mathrm{r}2}} = \braket{t|\Psi_\rho} \), with \( \mathcal{L}_\rho \) as in~\eqref{diff_t_space}, subject to \( \mathcal{B}[\braket{t|\Psi_\rho^{\mathrm{r}2}}] = 0 \). The general solution is
\begin{equation}
  \braket{t|\Psi_\rho^{\mathrm{r}2}} = \left( \beta - i \log t \right) \braket{t|\Psi_\rho}, \qquad \beta \in \mathbb{C} \, .
\end{equation}

Its position-space boundary value is
\begin{equation}
  \left. \braket{x|\Psi_\rho^{\mathrm{r}2}} \right|_{x=0} = \beta\, \mathfrak{z}(\rho) - i \mathfrak{z}^{(1)}(\rho) = -i \, \Gamma(\rho+1) \, (\rho-1) \, \zeta^{(1)}(\rho) \, .
\end{equation}
Since \( \mathfrak{z}(\rho)=0 \), the free constant drops out of the boundary value, and the Dirichlet condition reduces to \( \zeta^{(1)}(\rho)=0 \). The endpoint estimates for \( \braket{t|\Psi_\rho} \) show that the logarithmic factor preserves membership in \( \mathcal{D}(\hat{D})\cap\mathcal{D}(\hat{T}) \) and makes \( \mathcal{B}[\braket{t|\Psi_\rho^{\mathrm{r}2}}] \) vanish. The Dirichlet condition is therefore the only remaining requirement, so \( \hat{\mathcal{R}} \) admits a rank-\( 2 \) Jordan chain at \( i(\rho-1/2) \) precisely when \( \rho \) is a multiple zero.
\end{remark}

\section{Operator-Theoretic Formulation of RH}
\label{sec_RH}

The two eigenstate families of Sec.~\ref{subsec_adjoint} span the subspaces
\begin{equation}
\label{H_subspaces}
  \mathcal{H}_\Psi := \mathrm{span}\{\ket{\Psi_\rho}\}_{\rho\in\mathcal{Z}_\zeta}, \qquad \mathcal{H}_\Phi := \mathrm{span}\{\ket{\Phi_\rho}\}_{\rho\in\mathcal{Z}_\zeta} \, ,
\end{equation}
where \( \mathrm{span} \) denotes finite linear combinations. Since eigenstates with distinct eigenvalues are linearly independent, each family forms an algebraic basis of its subspace.

\begin{lemma}[Intertwiner \( \hat{W} \)]
\label{int_W}
There exists a linear operator
\begin{equation}
\label{W_op}
  \hat{W}:=\sum_{\rho\in\mathcal{Z}_\zeta}\lambda_\rho\ket{\Psi_\rho}\bra{\Psi_{1-\bar{\rho}}}\colon\mathcal{H}_\Phi\to\mathcal{H}_\Psi, \qquad \lambda_\rho \in \mathbb{C}^\times \, ,
\end{equation}
that intertwines \( \hat{\mathcal{R}} \) with its Hilbert-space adjoint \( \hat{\mathcal{R}}^* \) on \( \mathcal{H}_\Phi \):
\begin{equation}
\label{intertwining_rel}
  \hat{\mathcal{R}}\,\hat{W}\ket{U}=\hat{W}\,\hat{\mathcal{R}}^*\ket{U} \, , \qquad \ket{U}\in\mathcal{H}_\Phi \, .
\end{equation}
\end{lemma}

\begin{proof}
By Definition~\ref{def_zeros}, the involution \( \rho\mapsto1-\bar{\rho} \) maps \( \mathcal{Z}_\zeta \) bijectively to itself. The biorthogonality relation~\eqref{eq_biorthogonality} gives
\begin{equation}
  \hat{W}\ket{\Phi_\rho}=\lambda_{1-\bar{\rho}}\,\overline{N_\rho}\ket{\Psi_{1-\bar{\rho}}} \, .
\end{equation}
Thus only finitely many terms contribute on any element of \( \mathcal{H}_\Phi \), and \( \hat{W} \) is well defined with range contained in \( \mathcal{H}_\Psi \).

For the intertwining relation, Theorems~\ref{theo_eigen_R} and~\ref{dual_eigensystem} give \( \mathcal{H}_\Psi\subset\mathcal{D}(\hat{\mathcal{R}}) \), \( \mathcal{H}_\Phi\subset\mathcal{D}(\hat{\mathcal{R}}^*) \), and \( \hat{\mathcal{R}}^*\mathcal{H}_\Phi\subseteq\mathcal{H}_\Phi \). Hence both sides of~\eqref{intertwining_rel} are defined on \( \mathcal{H}_\Phi \). Since \( i((1-\bar{\rho})-1/2)=-i(\bar{\rho}-1/2) \), their actions on each basis vector agree:
\begin{equation}
  \hat{\mathcal{R}}\,\hat{W}\ket{\Phi_\rho} =-i\left(\bar{\rho}-1/2\right)\lambda_{1-\bar{\rho}}\,\overline{N_\rho}\ket{\Psi_{1-\bar{\rho}}} =\hat{W}\,\hat{\mathcal{R}}^*\ket{\Phi_\rho} \, .
\end{equation}
The identity therefore holds throughout \( \mathcal{H}_\Phi \) by linearity.
\end{proof}

\begin{remark}
\label{rem_W_as_V}
The intertwiner \( \hat{W} \) admits a formal analogue of the factorization~\eqref{V_general}. In the eigenbasis \( \{\ket{t}\}_{t>0} \), the generalized eigenvectors~\eqref{D_eigenbasis} of \( \hat{D} \) read \( \ket{\nu}=\int_0^\infty\braket{t|\nu}\ket{t}\,dt \), with \( \braket{t|\nu} \) given by~\eqref{nu_t_overlap}. Dividing the spectral coefficients of \( \ket{\Psi_\rho} \) by \( \omega(t) \) and analytically continuing~\eqref{nu_t_overlap} gives
\begin{equation}
\label{psi_to_nu}
  \omega(\hat{T})^{-1}\ket{\Psi_\rho}\overset{\mathrm{formal}}{=}\int_0^\infty t^{\rho-1}\ket{t}\,dt=\sqrt{2\pi}\,\frac{\Gamma(\rho)}{\Gamma(1-\rho)}\ket{i(\rho-1/2)} \, .
\end{equation}
Since \( t^{\rho-1}\notin L^2((0,\infty),dt) \), the state \( \ket{\Psi_\rho} \) lies outside the domain of the Hilbert-space inverse \( \omega(\hat{T})^{-1} \); the first equality therefore denotes only division of spectral coefficients. The parameter \( i(\rho-1/2) \) is real precisely when \( \rho \) lies on the critical line. There the ket belongs to the real-parameter generalized spectral family of \( \hat{D} \); off the line, it is understood through analytic continuation.

Under \( \rho\mapsto1-\bar{\rho} \), the gamma ratio becomes the reciprocal of its complex conjugate. Passing to the corresponding bra therefore cancels the gamma factors, yielding
\begin{equation}
  \hat{W}=\omega(\hat{T})\left[2\pi\sum_{\rho\in\mathcal{Z}_\zeta}\lambda_\rho\ket{i(\rho-1/2)}\bra{i(1/2-\bar{\rho})}\right]\omega(\hat{T}) \, .
\end{equation}
The middle factor commutes formally with the dilation differential expression, since the ket and bra in each summand have the same right and left eigenvalue \( i(\rho-1/2) \), respectively. For arbitrary complex coefficients \( \lambda_\rho \), this factor need not be formally symmetric. On the critical line it involves distributional spectral weights, while off the line it uses analytically continued eigenfunctions. This is therefore a formal spectral representation rather than a Borel functional-calculus identity.
\end{remark}

\begin{theorem}[Positivity, RH and Simplicity]
\label{RH_theo}
Consider \( \hat{W} \) in~\eqref{W_op}, with coefficients chosen so that \( \lambda_\rho>0 \) at every simple zero on the critical line. Then \( \hat{W}>0 \)---that is, \( \braket{\varphi|\hat{W}\,\varphi}>0 \) for every nonzero \( \ket{\varphi}\in\mathcal{H}_\Phi \)---if and only if every nontrivial Riemann zero is simple and lies on the critical line:
\begin{equation}
  \zeta^{(1)}(\rho)\neq0 \quad \text{and} \quad \Re(\rho)=1/2 \qquad \text{for all } \rho\in\mathcal{Z}_\zeta \, .
\end{equation}
\end{theorem}

\begin{proof}
Let \( \ket{\varphi}=\sum_{\rho\in\mathcal{Z}_\zeta}c_\rho\ket{\Phi_\rho} \) be a nonzero element of \( \mathcal{H}_\Phi \), with only finitely many nonzero coefficients. By~\eqref{eq_biorthogonality}, \( \braket{\varphi|\Psi_\rho}=\overline{c_\rho}\,N_\rho \), so the quadratic form is
\begin{equation}
\label{quadratic_form}
  \braket{\varphi|\hat{W}\,\varphi} =\sum_{\rho\in\mathcal{Z}_\zeta}\lambda_\rho\braket{\varphi|\Psi_\rho}\braket{\Psi_{1-\bar{\rho}}|\varphi} =\sum_{\rho\in\mathcal{Z}_\zeta}\lambda_\rho N_\rho\,\overline{N_{1-\bar{\rho}}}\,\overline{c_\rho}\,c_{1-\bar{\rho}} \, .
\end{equation}

Assume \( \hat{W}>0 \). For any \( \rho\in\mathcal{Z}_\zeta \), taking \( \ket{\varphi}=\ket{\Phi_\rho} \) gives
\begin{equation}
  \braket{\Phi_\rho|\hat{W}\,\Phi_\rho} =\lambda_\rho|N_\rho|^2\,\delta_{\rho,\,1-\bar{\rho}} \, .
\end{equation}
Since \( \ket{\Phi_\rho}\neq0 \), strict positivity forces \( 1-\bar{\rho}=\rho \) and \( N_\rho\neq0 \). Thus every nontrivial zero lies on the critical line and, by~\eqref{N_def}, is simple.

Conversely, suppose every nontrivial zero is simple and lies on the critical line. Then \( 1-\bar{\rho}=\rho \) and \( N_\rho\neq0 \) for every \( \rho\in\mathcal{Z}_\zeta \), while the choice of coefficients gives \( \lambda_\rho>0 \). Equation~\eqref{quadratic_form} therefore reduces to
\begin{equation}
  \braket{\varphi|\hat{W}\,\varphi}=\sum_{\rho\in\mathcal{Z}_\zeta}\lambda_\rho|N_\rho|^2\,|c_\rho|^2>0 \, ,
\end{equation}
since at least one coefficient \( c_\rho \) is nonzero.
\end{proof}

\begin{observation}[Weil--Bombieri Positivity Criterion]
\label{rem_weil_positivity}
In Bombieri's refinement of Weil's positivity criterion~\cite{weil1952formules,Bombieri2000}, RH is equivalent to strict positivity:
\begin{equation}
\label{weil_bombieri}
  Q(g):=\sum_{\rho}g(\rho)\,\overline{g(1-\bar{\rho})}>0 \, ,
\end{equation}
for every nonzero \( \tilde{g}\in\mathcal{C} \), where \( g(s)=\int_0^\infty\tilde{g}(t)\,t^{s-1}\,dt \) is its Mellin transform and the sum runs over the nontrivial zeros counted with multiplicity.

Our framework recovers a weighted version of the same reflection-paired structure. The pairings
\begin{equation}
  \braket{\varphi|\Psi_\rho}=\int_0^\infty t^{\rho-1}\omega(t)\braket{\varphi|t}\,dt=g(\rho) \, , \qquad \braket{\Psi_{1-\bar{\rho}}|\varphi}=\overline{g(1-\bar{\rho})} \, ,
\end{equation}
give the identification \( \tilde{g}(t)=\omega(t)\braket{\varphi|t} \) for states whose profile \( \tilde{g} \) lies in \( \mathcal{C} \). Conversely, every \( \tilde{g}\in\mathcal{C} \) arises from such a state by taking \( \braket{t|\varphi}=\overline{\tilde{g}(t)}/\omega(t) \), which is square-integrable because \( \omega \) is smooth and positive on the compact support of \( \tilde{g} \).

These test states need not lie in \( \mathcal{H}_\Phi \). When the coefficients \( \lambda_\rho \) have at most polynomial growth along the zeros, the series~\eqref{W_op} nevertheless defines an extension of \( \hat{W} \) to the algebraic sum of \( \mathcal{H}_\Phi \) and the subspace of test states. Indeed, the rapid decay of \( g \) in vertical strips, the standard zero-counting estimate, and the uniform boundedness of \( \|\Psi_\rho\| \) for \( 0<\Re(\rho)<1 \) give \( \sum_{\rho\in\mathcal{Z}_\zeta}|\lambda_\rho|\,\|\Psi_\rho\|\,|g(1-\bar{\rho})|<\infty \). Thus the defining vector series converges absolutely in \( \mathcal{H} \).

On these test states, the quadratic form becomes
\begin{equation}
  \braket{\varphi|\hat{W}\,\varphi} =\sum_{\rho\in\mathcal{Z}_\zeta}\lambda_\rho g(\rho)\,\overline{g(1-\bar{\rho})}=:Q_\lambda(g) \, .
\end{equation}
Choosing \( \lambda_\rho \) equal to the multiplicity of \( \rho \) recovers \( Q(g) \) exactly, since~\eqref{W_op} sums over distinct zeros. In particular, \( \lambda_\rho=1 \) recovers the classical form when all zeros are simple. For general coefficients, \( Q_\lambda \) retains the same reflection pairing but need not coincide with \( Q \).

The positivity criterion of Theorem~\ref{RH_theo} is therefore a weighted operator-theoretic counterpart of the Weil--Bombieri criterion. The domains remain essential: strict positivity of the classical form on Mellin test functions is equivalent to RH alone, whereas strict positivity of \( \hat{W} \) on \( \mathcal{H}_\Phi \), with the coefficients chosen as in Theorem~\ref{RH_theo}, is equivalent to RH together with simplicity.
\end{observation}

\begin{remark}
\label{rem_hilbert_polya}
A Hilbert--Pólya realization based on \( \hat{\mathcal{R}} \) would naturally be sought through the formal similarity transform
\begin{equation}
  \hat{H}:=\hat{W}^{-1/2}\,\hat{\mathcal{R}}\,\hat{W}^{1/2} \, .
\end{equation}
If all nontrivial zeros are simple, setting \( \kappa_\rho=1/\mathfrak{z}^{(1)}(\bar{\rho}) \) gives \( N_\rho=1 \). The biorthogonality relation~\eqref{eq_biorthogonality} then yields \( \hat{W}\ket{\Phi_{1-\bar{\rho}}}=\lambda_\rho\ket{\Psi_\rho} \). Since \( \lambda_\rho\neq0 \), the algebraic inverse exists and is given by
\begin{equation}
  \hat{W}^{-1}=\sum_{\rho\in\mathcal{Z}_\zeta}\frac{1}{\lambda_\rho}\ket{\Phi_{1-\bar{\rho}}}\bra{\Phi_\rho}\colon\mathcal{H}_\Psi\to\mathcal{H}_\Phi \, .
\end{equation}

To define \( \hat{W}^{\pm1/2} \) as positive self-adjoint operators, however, one needs an injective nonnegative self-adjoint realization of \( \hat{W} \). Its inverse and both powers are then defined by spectral calculus on their respective domains; the domain of the proposed product and its self-adjointness still require justification. Such a realization would make \( \hat{W} \) strictly positive on \( \mathcal{H}_\Phi \), already entailing RH by Theorem~\ref{RH_theo}. The same conclusion follows from strict positivity of the algebraic inverse on \( \mathcal{H}_\Psi \). Thus the positivity required for this transform already contains the desired zero-location conclusion, before self-adjointness of \( \hat{H} \) is established.

Independently, the full Riemann operator \( \hat{\mathcal{R}} \) on \( \mathcal{H} \) cannot become self-adjoint through a bounded invertible similarity, since its nonreal residual spectrum in Remark~\ref{residual_remark} would be preserved. A Hilbert--Pólya realization would therefore require a restriction to a suitable invariant subspace, a different Hilbert-space realization, or a transform outside this bounded invertible setting.
\end{remark}

\section{Outlook}
\label{sec_outlook}

Before closing, we comment on possible extensions of the framework to other \( L \)-functions (see, e.g., Ref.~\cite{lmfdb}). The completed zeta function \( \mathfrak{z}(s) \) arises as the boundary value \( \Psi_s(0) \) of the formal eigenfunction family of the Riemann operator \( \hat{\mathcal{R}} \) through the Mellin transform; the Dirichlet condition selects its zeros. The same construction therefore suggests extensions to suitable completions of other \( L \)-functions admitting Mellin representations with smooth nonvanishing weights, appropriate endpoint behavior, and the operator-domain properties required above.

Formulating the Generalized Riemann Hypothesis (GRH)~\cite{conrey2003,davenport2013multiplicative} as a strict positivity condition requires more. The intertwiner \( \hat{W} \) in~\eqref{W_op} requires the zero set to be invariant under \( \rho\mapsto1-\bar{\rho} \), as ensured by the conjugate functional equation after normalization to the central line \( \Re(s)=1/2 \). For extensions admitting the analogous eigensystem and biorthogonality relation, consider the intertwiner \( \hat{W} \) with coefficients \( \lambda_\rho\in\mathbb C^\times \) chosen positive at every simple zero on the critical line. The argument of Theorem~\ref{RH_theo} then carries over: strict positivity of \( \hat{W} \) is equivalent to GRH for the function in question together with simplicity of its nontrivial zeros. However, the functional equation supplies the required zero symmetry without establishing strict positivity. There are Dirichlet series that satisfy a functional equation yet violate the analogue of RH: suitable linear combinations of Dirichlet \( L \)-functions, such as the Davenport--Heilbronn function~\cite{davenport1936zeros1,davenport1936zeros2}, have zeros off the critical line. Whenever the analogous operator construction applies to such zeros, they rule out strict positivity of \( \hat{W} \). Establishing strict positivity therefore requires additional input beyond the Mellin representation and the functional equation. For arithmetic \( L \)-functions, the Euler product, which these counterexamples lack, is a natural source of such additional structure.

\section*{Acknowledgments}

The author is deeply grateful to Michael Berry, Felix Finster, Sebastian Hoerhold, Jon Keating, Klaus Richter, Jonas Ringo, and Juan Diego Urbina for many insightful discussions.






\bibliographystyle{unsrt}
\bibliography{../../bib_zeta.bib}

\begin{thebibliography}{10}

\bibitem{riemann1859ueber}
Bernhard Riemann.
\newblock Ueber die anzahl der primzahlen unter einer gegebenen grosse.
\newblock {\em Ges. Math. Werke und Wissenschaftlicher Nachla{\ss}}, 2(145-155):2, 1859.

\bibitem{edwards1974riemannaes}
Harold~M Edwards.
\newblock {\em Riemann's Zeta Function}, volume~58.
\newblock Academic press, 1974.

\bibitem{titchmarsh1986theory}
Edward~Charles Titchmarsh and David~Rodney Heath-Brown.
\newblock {\em The theory of the Riemann zeta-function}.
\newblock Oxford university press, 1986.

\bibitem{okubo1998lorentz}
Susumu Okubo.
\newblock Lorentz-invariant hamiltonian and riemann hypothesis.
\newblock {\em Journal of Physics A: Mathematical and General}, 31(3):1049, 1998.

\bibitem{connes1999trace}
Alain Connes.
\newblock Trace formula in noncommutative geometry and the zeros of the riemann zeta function.
\newblock {\em Selecta Mathematica}, 5(1):29--106, 1999.

\bibitem{kuipers2014quantum}
Jack Kuipers, Quirin Hummel, and Klaus Richter.
\newblock Quantum graphs whose spectra mimic the zeros of the riemann zeta function.
\newblock {\em Physical Review Letters}, 112(7):070406, 2014.

\bibitem{berry1999}
Michael~V Berry and Jonathan~P Keating.
\newblock H= xp and the riemann zeros.
\newblock In {\em Supersymmetry and Trace Formulae}, pages 355--367. Springer, 1999.

\bibitem{bender2017hamiltonian}
Carl~M Bender, Dorje~C Brody, and Markus~P M{\"u}ller.
\newblock Hamiltonian for the zeros of the riemann zeta function.
\newblock {\em Physical Review Letters}, 118(13):130201, 2017.

\bibitem{yakaboylu2024hamiltonian}
Enderalp Yakaboylu.
\newblock Hamiltonian for the hilbert--p{\'o}lya conjecture.
\newblock {\em Journal of Physics A: Mathematical and Theoretical}, 57(23):235204, 2024.

\bibitem{weil1952formules}
André Weil.
\newblock Sur les ``formules explicites'' de la théorie des nombres premiers.
\newblock {\em Comm. Sem. Math. Univ. Lund (Supplement dédié à Marcel Riesz)}, pages 252--265, 1952.

\bibitem{Bombieri2000}
Enrico Bombieri.
\newblock Remarks on weil's quadratic functional in the theory of prime numbers, i.
\newblock {\em Rendiconti Lincei. Matematica e Applicazioni}, 11(3):183--233, 2000.

\bibitem{endres2010berry}
Sebastian Endres and Frank Steiner.
\newblock The berry-keating operator on and on compact quantum graphs with general self-adjoint realizations.
\newblock {\em Journal of Physics A: Mathematical and Theoretical}, 43(9):095204, 2010.

\bibitem{reed1972methods}
Michael Reed and Barry Simon.
\newblock {\em Methods of Modern Mathematical Physics, Vol 1 and 2.}
\newblock New York, London: Academic Press, 1972.

\bibitem{teschl2014mathematical}
Gerald Teschl.
\newblock {\em Mathematical Methods in Quantum Mechanics}, volume 157.
\newblock American Mathematical Soc., 2014.

\bibitem{lmfdb}
The {LMFDB Collaboration}.
\newblock The {L}-functions and modular forms database.
\newblock \url{https://www.lmfdb.org}, 2025.
\newblock [Online; accessed 10 November 2025].

\bibitem{conrey2003}
J.~B. Conrey.
\newblock The riemann hypothesis.
\newblock {\em Notices of the American Mathematical Society}, 50(3):341--353, 2003.

\bibitem{davenport2013multiplicative}
Harold Davenport.
\newblock {\em Multiplicative number theory}, volume~74.
\newblock Springer Science \& Business Media, 2013.

\bibitem{davenport1936zeros1}
H.~Davenport and H.~Heilbronn.
\newblock On the zeros of certain {D}irichlet series.
\newblock {\em J. London Math. Soc.}, 11:181--185, 1936.

\bibitem{davenport1936zeros2}
H.~Davenport and H.~Heilbronn.
\newblock On the zeros of certain {D}irichlet series, second paper.
\newblock {\em J. London Math. Soc.}, 11:307--312, 1936.

\end{thebibliography}

\end{document}